\documentclass{elsart}
\usepackage{graphicx}
\pdfoutput = 1

\newcommand{\Oh}[1]
    {\ensuremath{\mathcal{O} \hspace{-.5ex} \left( {#1} \right)}}
\newcommand{\occ}[2]
    {\ensuremath{\mathrm{occ} \left( {#1}, {#2} \right)}}

\begin{document}

\begin{frontmatter}

\title{Tight Bounds for Online Stable Sorting}

\author{Travis Gagie}
\address{Department of Computer Science\\
    University of Eastern Piedmont, Italy}
\ead{travis@mfn.unipmn.it}

\author{Yakov Nekrich}
\address{Department of Computer Science\\
    University of Bonn, Germany}
\ead{yasha@cs.uni-bonn.com}

\begin{abstract}
Although many authors have considered how many ternary comparisons it takes to sort a multiset $S$ of size $n$, the best known upper and lower bounds still differ by a term linear in $n$.  In this paper we restrict our attention to online stable sorting and prove upper and lower bounds that are within \(o (n)\) not only of each other but also of the best known upper bound for offline sorting.  Specifically, we first prove that if the number of distinct elements \(\sigma = o (n / \log n)\), then \((H + 1) n + o (n)\) comparisons are sufficient, where $H$ is the entropy of the distribution of the elements in $S$.  We then give a simple proof that \((H + 1) n - o (n)\) comparisons are necessary in the worst case.
\end{abstract}

\begin{keyword}
Analysis of algorithms; online algorithms; sorting.
\end{keyword}

\end{frontmatter}

\section{Introduction} \label{sec:intro}

Comparison-based sorting is perhaps the most studied problem in computer science, but there remain basic open questions about it.  For example, how many ternary comparisons are needed to sort a multiset $S$ of size $n$?  By a ternary comparison, we mean one that can return $<$, $=$ or $>$; we count only comparisons between elements of the multiset, not those between data generated by the algorithm.  Over thirty years ago, Munro and Spira~\cite{MS76} proved distribution-sensitive upper and lower bounds that differ by $\Oh{n \log \log \sigma}$, where $\sigma$ is the number of distinct elements in $S$.  Their bounds have been improved in a series of papers --- summarized in Table~\ref{tab:bounds} --- and now the best known upper and lower bounds differ by a term linear in $n$ (about \((1 + \log e) n \approx 2.44 n\) when \(\sigma = o (n)\)).  In this paper we restrict our attention to online stable sorting and prove distribution-sensitive upper and lower bounds that are within \(o (n)\) not only of each other but also of the best known upper bound for offline sorting.  When \(\sigma = o (n / \log n)\) we can now say, for example, exactly how many comparisons on average per element are needed to sort online and stably as $n$ goes to infinity.

\begin{table}[t]
\caption{Bounds on the number of ternary comparisons needed to sort a multiset offline.}
\label{tab:bounds}
\resizebox{\textwidth}{!}
{\begin{tabular}{r|cc}
& upper bound & lower bound\\
\hline \rule{0ex}{3ex}
Munro and Raman~\cite{MR91} && \((H - \log e) n + \Oh{\log n}\)\\[1ex]
Fischer~\cite{Fis84} & \((H + 1) n - \sigma\) & \((H - \log H) n - \Oh{n}\)\\[1ex]
Dobkin and Munro~\cite{DM80} && \(\left( H - n \log \left( \log n - \frac{\sum_i \occ{a_i}{S} \log \occ{a_i}{S}}{n} \right) \right) n - \Oh{n}\)\\[1ex]
Munro and Spira~\cite{MS76} & \(n H + \Oh{n}\) & \(n H - (n - \sigma) \log \log \sigma - \Oh{n}\)
\end{tabular}}
\end{table}

Stable sorting algorithms preserve the order of equal elements, and we define online sorting algorithms to be those that process $S$ element by element and keep those already seen in sorted order.  We believe our definition is reasonable because, if we want a sorting algorithm to be stable and each comparison it makes to involve the most recently read element --- to prevent it delaying all its decisions until it has read all of $S$ --- then it must determine each element's rank and distinctness among those already seen, before reading the next element.  To see why, suppose the most recently read element is $x$ and, for some element $y$ already seen, the algorithm reads the next element without determining whether \(x < y\), \(x = y\), or \(x > y\).  Without loss of generality, assume \(x \leq y\).  If all the remaining elements are either strictly less than $x$ or strictly greater than $y$, then the algorithm has no further opportunity to determine $x$ and $y$'s relative order.  Notice that, since the algorithm has no information about an element until it has been involved in a comparison, it makes no difference in the worst case whether the algorithm chooses the order in which the elements are to be read, or an adversary does.

Splay-sort, for example --- sorting by insertions into a splay-tree~\cite{ST85} --- is both online and stable.  It uses $\Oh{(H + 1) n}$ comparisons and time, where \(H = \sum_{i = 1}^\sigma \frac{\occ{a_i}{S}}{n} \log \frac{n}{\occ{a_i}{S}}\) is the entropy of the distribution of the elements, \(a_1, \ldots, a_i\) are the distinct elements, $\occ{a_i}{S}$ is the number times $a_i$ occurs in $S$, and $\log$ means $\log_2$.  Such bounds are called distribution-sensitive because they depend not only on $n$ but also on the distribution of elements.  It is not hard to show splay-sort's bound is within a constant factor of the optimal offline bound: we can encode $S$ by recording the result of each comparison in a constant number of bits, together with an $\Oh{\sigma \log n}$-bit table containing the distinct elements' frequencies; we generally cannot encode $S$ in fewer than \(n H\) bits; we must also use at least one comparison per element; therefore, we must use \(\Omega ((H + 1) n)\) comparisons.  Notice that, since we can implement a ternary comparison as two binary comparisons --- which can return only $\leq$ or $>$ --- and asymptotic notation can hide a two-fold increase, splay-sort's bound holds for both types of comparison.  In contrast, our results in this paper show that ternary comparisons are slightly more powerful.

Our starting point is another online stable sorting algorithm, due to Fredman~\cite{Fre76} and based on a result by Gilbert and Moore~\cite{GM59}.  In Section~\ref{sec:binary ub} we show how to use this algorithm to sort $S$ online and stably using at most \((H + 2) n + o (n)\) binary comparisons and $\Oh{\log \sigma}$ time per comparison when \(\sigma = o (n / \log n)\), or constant time per comparison when \(\sigma = o \left( \sqrt{n / \log n} \right)\).  In Section~\ref{sec:binary lb} we give a simple proof that \((H + 2) n - o (n)\) binary comparisons are necessary in the worst case.  In Section~\ref{sec:ternary ub} we turn our attention to sorting with ternary comparisons and show how to modify Fredman's algorithm to use at most \((H + 1) n + o (n)\) ternary comparisons when \(\sigma = o (n / \log n)\), by replacing Gilbert and Moore's result with one by Mehlhorn~\cite{Meh77}; the time bounds remain the same.  In Section~\ref{sec:ternary lb} we modify our lower bound to show \((H + 1) n - o (n)\) ternary comparisons are necessary in the worst case.  Notice our upper and lower bounds here are within \(o (n)\) not only of each other but also of the best known upper bound for offline sorting, \((H + 1) n - \sigma\), due to Fischer~\cite{Fis84} and shown in the second row of Table~\ref{tab:bounds}.  We hope studying online sorting will ultimately provide insights that will help narrow the gap between the upper and lower bounds for offline sorting.  Since sorting is fundamental, studying online sorting may even provide insights into online versions of other problems.  We must admit, the main insights we use in this paper came not from studying the literature about sorting --- although they easily could have --- but from writing a recent paper~\cite{GN09} about adaptive prefix coding.

\section{Upper bound on binary comparisons} \label{sec:binary ub}

Fredman~\cite{Fre76} showed how, given an $n$-tuple \((x_1, \ldots, x_n)\) drawn from a universe $U$, we can determine its components one by one using a total of at most \(\log |U| + 2 n\) binary comparisons.  To determine $x_i$, Fredman's algorithm restricts its attention to the candidate $n$-tuples consistent with \(x_1, \ldots, x_{i - 1}\); counts how often each value occurs as the $i$th component in these candidates; uses the following theorem by Gilbert and Moore~\cite{GM59} to build a leaf-oriented binary search tree for the values' normalized frequencies; and inserts $x_i$ into that tree.  The key observation is that, if the algorithm uses $c$ comparisons to determine $x_i$, then fewer than a \((1 / 2^{c - 2})\)-fraction of the candidates consistent with \(x_1, \ldots, x_{i - 1}\) are also consistent with $x_i$.

\begin{thm}[Gilbert and Moore, 1959] \label{thm:GM59}
Given a probability distribution \(p_1, \ldots, p_k\) containing only positive probabilities, we can build an ordered binary tree whose leaves are at depths at most \(\lceil \log (1 / p_1) \rceil + 1, \ldots, \lceil \log (1 / p_k) \rceil + 1\).
\end{thm}

\noindent Fredman's result immediately implies that, given the frequencies of the distinct elements in $S$, we can sort it online and stably using at most
\[\log \left( \rule{0ex}{2ex} \occ{a_1}{S}, \ldots, \occ{a_\sigma}{S} \right) ! + 2 n\]
comparisons, where \(\left( \occ{a_1}{S}, \ldots, \occ{a_\sigma}{S} \right) ! = n! / \prod_{i = 1}^\sigma \occ{a_i}{S}!\) is the number of strings in which each element $a_i$ appears $\occ{a_i}{S}$ times.  By the following lemma, this bound is at most \((H + 2) n + o (n)\).

\begin{lem} \label{lem:multinomial} \(\log \left( \rule{0ex}{2ex} \occ{a_1}{S}, \ldots, \occ{a_\sigma}{S} \right)! \leq n H + \Oh{\log n}\).
\end{lem}

\begin{pf}
Robbins' extension~\cite{Rob55} of Stirling's Formula,
\[\sqrt{2 \pi} x^{x + 1 / 2} e^{-x + 1 / (12 x + 1)}
< x!
< \sqrt{2 \pi} x^{x + 1 / 2} e^{-x + 1 / (12 x)}\,,\]
implies
\[x \log x - x \log e
< \log x!
\leq x \log x - x \log e + \Oh{\log x}\,.\]
Therefore, since \(\sum_i \occ{a_i}{S} = n\), straightforward calculation shows that
\[\log \left( \rule{0ex}{2ex} \occ{a_1}{S}, \ldots, \occ{a_\sigma}{S} \right) !
= \log n! - \sum_{i = 1}^\sigma \log \occ{a_i}{S}!\]
is at least \(n H - \Oh{\sigma \log (n / \sigma)}\) and at most \(n H + \Oh{\log n}\). \qed
\end{pf}

\noindent Of course, given the frequencies of the distinct elements in $S$, we can use Theorem~\ref{thm:GM59} to build a single leaf-oriented binary search tree whose leaves store the distinct elements in $S$, with each distinct element $a_j$ at depth at most \(\left\lceil \rule{0ex}{2ex} \log (n / \occ{a_j}{S}) \right\rceil\); we can then insert each element into that tree using a total of at most \((H + 2) n + o (n)\) comparisons.  We are interested in Fredman's algorithm because we can use it to sort $S$ using \((H + 2) n + o (n)\) comparisons even when we are not given the frequencies in advance, as long as \(\sigma = o (n / \log n)\).  To prove this, we start by modifying Theorem~\ref{thm:GM59}.

\begin{lem} \label{lem:GM59}
Given a probability distribution \(p_1, \ldots, p_k\) containing only positive probabilities, in $\Oh{k}$ time we can build an ordered binary tree on \(2 k + 1\) leaves with height at most \(\left\lceil \rule{0ex}{2ex} \log (1 / \min_{1 \leq i \leq k} \{p_i\}) \right\rceil + 1\) whose even-numbered leaves are at depths at most \(\lceil \log (1 / p_1) \rceil + 1, \ldots, \lceil \log (1 / p_k) \rceil + 1\).
\end{lem}

\begin{pf}
For \(1 \leq i \leq 2 k + 1\), consider the $i$th term in the sequence
\[0, \frac{p_1}{2}, p_1, \left( p_1 + \frac{p_2}{2} \right), (p_1 + p_2), \ldots, \left( \sum_{j = 1}^{k - 1} p_j + \frac{p_k}{2} \right), 1\,.\]
If $i$ is even, then the $i$th term differs from every other term by at least \(p_{i / 2} / 2\), so its binary representation $b_i$ differs from all of theirs in one of its first \(\left\lceil \log (1 / p_{i / 2}) \right\rceil + 1\) bits after the binary point.  Therefore, in a trie containing the shortest prefixes of \(b_1, \ldots, b_{2 k + 1}\) necessary to uniquely identify them, the even-numbered leaves are at depths at most \(\lceil \log (1 / p_1) \rceil + 1, \ldots, \lceil \log (1 / p_k) \rceil + 1\);
\linebreak
it follows that the odd-numbered leaves are all at depths at most
\linebreak
\(\left\lceil \rule{0ex}{2ex} \log (1 / \min_{1 \leq i \leq k} \{p_i\}) \right\rceil + 1\).  Since such a tree exists, we can use an algorithm by Kirkpatrick and Klawe~\cite{KK85} to build one in $\Oh{k}$ time. \qed
\end{pf}

\noindent We need not do anything to process the first element, \(S [1]\), since any singleton set is trivially sorted.  (For convenience, we write \(S [i]\) to indicate the $i$th element in $S$ and \(S [i..j]\) to indicate the $i$th through $j$th elements.)  Assume for the moment that we are not concerned with how much time we use to sort $S$.  To process \(S [i]\) for \(i \geq 2\), we first apply Lemma~\ref{lem:GM59} to the probability distribution \(\occ{a_1}{S [1..(i - 1)]} / (i - 1), \ldots, \occ{a_k}{S [1..(i - 1)]} / (i - 1)\), where \(a_1, \ldots, a_k\) are the distinct elements in \(S [1..(i - 1)]\).  Let $T_i$ be the resulting tree.  We label $T_i$'s even-numbered leaves with \(a_1, \ldots, a_k\) and its odd-numbered leaves with blanks; at each internal node, we store the labels of the rightmost leaf in that node's left subtree and the leftmost leaf in its right subtree, one of which is always a blank.  Finally, we insert \(S [i]\) into $T_i$ as follows: at each internal node that stores a blank and a key $a_j$, in that order, we check whether \(a_j \leq S [i]\) and descend to the right if so and to the left if not; at each internal node that stores a key $a_j$ and a blank, in that order, we check whether \(S [i] \leq a_j\) and descend to the left if so and to the right if not.  If \(S [i]\) is an occurrence of some $a_j$ that has already occurred, then we eventually reach the $j$th even-numbered leaf and, thus, use at most \(\lceil \log ((i - 1) / \occ{a_j}{S [1..(i - 1)]}) \rceil + 1\) comparisons.  On the other hand, if \(S [i]\) is not equal to any previous character and has rank $j$ in \(S [1..(i - 1)]\), then we eventually reach the $j$th odd-numbered leaf and, thus, use at most \(\lceil \log (i - 1) \rceil + 1\) comparisons.

\begin{thm} \label{thm:binary ub}
We can sort $S$ online and stably using at most \((H + 2) n + o (n)\) binary comparisons when \(\sigma = o (n / \log n)\).
\end{thm}

\begin{pf}
Since the multiset
\[\left\{ \rule{0ex}{2ex} \occ{S [i]}{S [1..(i - 1)]}\,:\,1 \leq i \leq n \right\}
= \bigcup_{j = 1}^\sigma \{0, \ldots, \occ{a_\sigma}{S} - 1\}\,,\]
we use at most
\begin{eqnarray*}
\lefteqn{\sum_{i = 1}^n \left( \left\lceil \log \frac{i - 1}
    {\max \left( \rule{0ex}{2ex} \occ{S [i]}{S [1..(i - 1)]}, 1 \right)} \right\rceil + 1 \right)}\\
& < & \sum_{i = 1}^n \log (i - 1) -
    \sum_{i = 1}^n \log \max \left( \rule{0ex}{2ex} \occ{S [i]}{S [1..(i - 1)]}, 1 \right) + 2 n\\
& = & \log (n - 1)! - \sum_{j = 1}^\sigma \log (\occ{a_j}{S} - 1)! + 2 n\\
& \leq & \log n! - \sum_{j = 1}^\sigma \log \occ{a_j}{S}! + 2 n + \Oh{\sigma \log n}\\
& = & \log \left( \rule{0ex}{2ex} \occ{a_1}{S}, \ldots, \occ{a_\sigma}{S} \right)! + 2 n + \Oh{\sigma \log n}
\end{eqnarray*}
comparisons.  By Lemma~\ref{lem:multinomial}, this is at most \((H + 2) n + \Oh{\sigma \log n}\) and, thus, \((H + 2) n + o (n)\) when \(\sigma = o (n / \log n)\). \qed
\end{pf}

\noindent To make our modification of Fredman's algorithm run in $\Oh{\log \sigma}$ time per comparison rather than $\Oh{\sigma}$ time, we do not actually build the trees \(T_2, \ldots, T_n\) but, instead, maintain an augmented AVL-tree~\cite{AL62} that stores the partial sums of the frequencies of the distinct elements seen so far.  Since each $T_i$ is defined by partial sums (recall the proof of Lemma~\ref{lem:GM59}), we can use this AVL-tree as an implicit representation of $T_i$ when processing \(S [i]\).  In particular, it is not difficult to use it to compute the labels stored at each internal node of $T_i$ in $\Oh{\log \sigma}$ time.

If \(\sigma = o \left( \sqrt{n / \log n} \right)\), on the other hand, then we can use $\Oh{1}$ time per comparison.  In this case, we build only a few of the trees \(T_2, \ldots, T_n\).  Specifically, we build a new tree whenever the number of elements processed since we built the last tree is equal to the number of distinct elements we had processed at that time; it follows that we need only $\Oh{1}$ amortized time per element.  Since we want the tree into which we insert \(S [i]\) always to contain all the distinct elements in \(S [1..(i - 1)]\), at each odd-numbered leaf we store a pointer to an AVL-tree.  If we reach an odd-numbered leaf when processing \(S [i]\), then we insert \(S [i]\) into the corresponding AVL-tree in the standard way, implementing each ternary comparison as two binary comparisons.  Calculation shows we use a total of at most
\begin{eqnarray*}
&& \hspace{-3ex} \sum_{i = 1}^n \left( \left\lceil \log \frac{i - 1}
    {\max \left( \rule{0ex}{2ex} \occ{S [i]}{S [1..(i - 1)]} - \sigma, 1 \right)} \right\rceil + 1 \right) +
    \Oh{\sigma^2 \log \sigma}\\[1ex]
& \leq & \log n!
    - \sum_{\occ{a_j}{S} > \sigma} \log \left( \rule{0ex}{2ex} \occ{a_j}{S} - \sigma - 1 \right)! +
    2 n + \Oh{\sigma^2 \log \sigma}\\
& \leq & \log n! - \sum_{j = 1}^\sigma \log \occ{a_j}{S}! + \sigma \log n^{\sigma + 1} +
    2 n + \Oh{\sigma^2 \log \sigma}\\
& \leq & (H + 2) n + \Oh{\sigma^2 \log n}
\end{eqnarray*}
comparisons and, thus, \((H + 2) n + o (n)\) when \(\sigma = o \left( \sqrt{n / \log n} \right)\).  Notice we subtract $\sigma$ from \(\occ{S [i]}{S [1..(i - 1)]}\) in the formula above because we always build the tree with which we process \(S [i]\) sometime after processing \(S [i - \sigma]\).

\section{Lower bound on binary comparisons} \label{sec:binary lb}

Consider any online stable sorting algorithm that uses only binary comparisons to determine the stably sorted order of the elements in $S$.  Since it is online, it must determine each element's rank relative to the distinct elements already seen before moving on to the next element.  Since it uses only binary comparisons, we can view its strategy for each element as a binary decision tree whose even-numbered leaves are labelled with the distinct elements already seen and whose odd-numbered leaves correspond to the intervals in which the next distinct element could lie.  (We consider its strategy as a decision tree rather than a search tree because we do not want to specify what comparisons it makes at internal nodes.)  If the current element has been seen before, then the algorithm reaches the even-numbered leaf labelled with that element in the decision tree; if not, then it reaches an odd-numbered leaf.  In both cases, the number of comparisons the algorithm uses is at least the depth of the leaf.

Suppose an adversary starts by presenting one copy of each of \(\sigma - 1\) distinct elements; after that, it always considers the algorithm's strategy for the current element as a binary decision tree with \(2 (\sigma - 1) + 1\) leaves, and presents the label of the deepest even-numbered leaf (except that it presents the $\sigma$th distinct element as the last element in $S$).  It is not difficult to show that, for the right choice of $\sigma$, the adversary thus forces the algorithm to use \((H + 2) n - o (n)\) comparisons.

\begin{thm} \label{thm:binary lb}
In the worst case we need to make at least \((H + 2) n - o (n)\) binary comparisons to sort $S$ online and stably.
\end{thm}

\begin{pf}
Suppose \(\sigma = 2^{\lfloor \log (n / \log n) \rfloor} + 1\).  Since $\sigma$ is 1 more than a power of 2, any binary tree on \(2 (\sigma - 1) + 1\) leaves has an even-numbered leaf (in fact, some consecutive pair of leaves) at depth at least
\[\log (\sigma - 1) + 2
> \log \sigma - \frac{\log e}{\sigma - 1} + 2
= \log \sigma + 2 - o (1)\,.\]
Therefore, since any distribution on $\sigma$ elements has entropy at most \(\log \sigma\), some even-numbered leaf has depth at least \(H + 2 - o (1)\).  It follows that the adversary forces the algorithm to use a total of at least \((H + 2 - o (1)) (n - \sigma) = (H + 2) n + o (n)\) comparisons to process the $\sigma$th through \((n - 1)\)st elements of $S$. \qed
\end{pf}

\noindent We have the adversary hold the $\sigma$th distinct element in reserve until the end because, this way, our lower bound holds even when the algorithm knows $\sigma$ in advance.  If the adversary started by presenting all $\sigma$ distinct elements then, since the algorithm would know it had already seen all the distinct elements, from then on it could use a decision tree on only $\sigma$ leaves.

\section{Upper bound on ternary comparisons} \label{sec:ternary ub}

Shortly after Munro and Spira~\cite{MS76} and Fredman~\cite{Fre76} published their results, Mehlhorn~\cite{Meh77} published a generalization of Gilbert and Moore's~\cite{GM59}.  In this paper we do not need Mehlhorn's full result, only the weaker version below.  We note that, although Mehlhorn's proof was not based on partial sums, Knuth~\cite{Knu78} quickly gave another that was.

\begin{thm}[Mehlhorn, 1977] \label{thm:Meh77}
Given a probability distribution \(p_1, \ldots, p_k\) containing only positive probabilities, in $\Oh{k}$ time we can build an ordered binary tree whose nodes, from left to right, are at depths at most \(\log (1 / p_1), \ldots,\)
\(\log (1 / p_k)\).
\end{thm}

\noindent Given the frequencies of the distinct elements in $S$, we can use Theorem~\ref{thm:Meh77} to build a (node-oriented) binary search tree whose nodes store the distinct elements in $S$, with each distinct element $a_j$ at depth at most \(\log (n / \occ{a_j}{S})\); we can then insert each element into that tree using a total of at most \((H + 1) n + o (n)\) comparisons.  Notice the first term in this bound is \((H + 1) n\) rather than \(n H\) because the number of comparisons used to reach and stop at an internal node is 1 more than its depth; the last comparison is performed at the internal node itself and indicates that we should stop there.  Because the tree contains all the distinct elements in $S$, however, we need never perform a comparison at a leaf.

We can sort $S$ using at most \((H + 1) n + o (n)\) comparisons even when we are not given the frequencies in advance, as long as \(\sigma = o (n / \log n)\), by modifying Fredman's algorithm to use Theorem~\ref{thm:Meh77} and ternary comparisons, instead of Theorem~\ref{thm:GM59} and binary comparisons.  Again, we need not do anything to process \(S [1]\).  Assume for the moment that we are again not concerned with how much time we use to sort $S$.  To process \(S [1]\) for \(i \geq 2\), we first apply Theorem~\ref{thm:Meh77} to the probability distribution \(\occ{a_1}{S [1..(i - 1)]} / (i - 1), \ldots, \occ{a_k}{S [1..(i - 1)]} / (i - 1)\), where \(a_1, \ldots, a_k\) are again the distinct elements in \(S [1..(i - 1)]\).  Let $T_i$ be the resulting tree.  We store \(a_1, \ldots, a_k\) at $T_i$'s nodes, from left to right, thus making it a binary search tree.  Finally, we insert \(S [i]\) into $T_i$ in the standard way, using ternary comparisons.  If \(S [i]\) is an occurrence of some $a_j$ that has already occurred, then we eventually reach and stop at the node storing $a_j$ and, thus, use at most \(\log (1 / p_j) + 1\) comparisons.  Notice that, because $T_i$ may not contain all the distinct elements in $S$, only those in \(S [1..(i - 1)]\), now we must also perform a comparison if we reach a leaf.  On the other hand, if \(S [i]\) is not equal to any previous character and has rank $j$ in \(S [1..(i - 1)]\), then we eventually reach a leaf storing either $a_{j - 1}$ or $a_j$ --- i.e., either the predecessor or the successor of \(S [i]\) in \(S [1..(i - 1)]\) --- and, thus, use at most \(\log (i - 1) + 1\) comparisons.  Essentially the same calculations as in the proof of Theorem~\ref{thm:binary ub} show we use a total of at most \((H + 1) n + o (n)\) comparisons.

\begin{thm} \label{thm:ternary ub}
We can sort $S$ online and stably using at most \((H + 1) n + o (n)\) ternary comparisons when \(\sigma = o (n / \log n)\).
\end{thm}

\noindent Because Knuth's proof of Theorem~\ref{thm:Meh77} is based on partial sums, we can reuse the techniques described in Section~\ref{sec:binary ub} to make this algorithm run in $\Oh{\log \sigma}$ time per comparison, or $\Oh{1}$ time per comparison when \(\sigma = \Oh{\sqrt{n / \log n}}\).

\section{Lower bound on ternary comparisons} \label{sec:ternary lb}

Now consider any online stable sorting algorithm that uses ternary comparisons to determine the stably sorted order of the elements in $S$.  Again, since it is online and stable, it must determine each element's rank relative to the distinct elements already seen before moving on to the next element.  Since it uses ternary comparisons, we can view its strategy for each element as an extended binary search tree whose internal nodes store the distinct elements already seen and whose leaves correspond to the intervals in which the next distinct element could lie.  If the current element has been seen before, then the algorithm reaches and stops at the internal node storing that element, and the number of comparisons that it uses is at least the depth of that internal node plus 1.  If not, then it reaches a leaf and the number of comparisons it uses is at least the depth of that leaf.

To see why we can consider the algorithm's strategy for each element as an extended binary search tree, first consider it as a ternary decision tree.  Since the algorithm keeps the elements already seen in sorted order and middle children are reached only when a comparison returns $=$, the subtrees rooted at those children are all degenerate.  Therefore, without loss of generality, we can assume that middle children are leaves and labelled with the distinct elements already seen, while leaves that are left or right children correspond to the intervals in which the next distinct element could lie.  By deleting each middle child and storing its label at its parent, we obtain an extended binary search tree whose internal nodes store the distinct elements already seen and whose leaves correspond to the intervals in which the next distinct element could lie.

Suppose an adversary again starts by presenting one copy of each of \(\sigma - 1\) distinct elements; after that, it always considers the algorithm's strategy for the current element as an extended binary search tree with \(\sigma - 1\) internal nodes, and presents the element stored at the deepest internal node (except that it presents the $\sigma$th distinct element as the last element in $S$).  Essentially the same argument as in the proof of Theorem~\ref{thm:binary ub} shows that, for the right choice of $\sigma$, the adversary thus forces the algorithm to use \((H + 1) n - o (n)\) comparisons.

\begin{thm} \label{thm:ternary lb}
In the worst case we need to make at least \((H + 1) n - o (n)\) ternary comparisons to sort $S$ online and stably.
\end{thm}

\begin{pf}
Suppose again that \(\sigma = 2^{\lfloor \log (n / \log n) \rfloor} + 1\).  Since $\sigma$ is 1 more than a power of 2, any binary search tree with \(\sigma - 1\) internal nodes has an internal node at depth at least \(\log (\sigma - 1)\).  Essentially the same calculations as in the proof of Theorem~\ref{thm:binary ub} show that \(\log (\sigma - 1) \geq H - o (1)\).  Therefore, when the algorithm is processing the $\sigma$th through \((n - 1)\)st elements in $S$, the adversary always chooses an internal node at depth at least \(H - o (1)\) and, so, forces the algorithm to use at least \(H + 1 - o (1)\) comparisons per element.  Thus, the algorithm uses a total of at least \((H + 1 - o (1)) (n - \sigma) = (H + 1) n - o (n)\) comparisons.
\qed
\end{pf}

\noindent We again have the adversary hold the $\sigma$th distinct element in reserve until the end, because, this way, our lower bound holds even when the algorithm knows $\sigma$ in advance.  If the adversary started by presenting all $\sigma$ distinct elements then, since the algorithm would know it had already seen all the distinct elements, it would not have to make a comparison at any internal node with two leaves as children, in particular the deepest internal node.

Combining Theorems~\ref{thm:ternary ub} and~\ref{thm:ternary lb} yields the following corollary which, for \(\sigma = o (n / \log n)\), tells us exactly how many comparisons are needed on average per element to sort online and stably as $n$ goes to infinity.

\begin{cor} \label{cor:ternary}
When \(\sigma = o (n / \log n)\), in the worst case it takes an average of \(H + 1 \pm o (1)\) ternary comparisons per element to sort $S$ online and stably.
\end{cor}

\bibliographystyle{plain}
\bibliography{online_sorting}

\end{document}